\documentclass[aps,amssymb,10pt,showpacs,showkeys,letterpaper,twocolumn]{revtex4}
\usepackage{graphicx}
\usepackage{amsmath}
\usepackage{epsfig}
\usepackage{bm}

\newcommand{\be}{\begin{equation}}
\newcommand{\ee}{\end{equation}}
\newcommand{\beq}{\begin{eqnarray}}
\newcommand{\eeq}{\end{eqnarray}}

\begin{document}

\title{About the coordinate time for photons in Lifshitz Space-times}

\author{ J. R. Villanueva }
 \email{jose.villanuevalob@uv.cl}
\affiliation{ Departamento de F\'{\i}sica y Astronom\'ia, Facultad
de Ciencias, Universidad de Valpara\'iso, Gran Breta\~na 1111,\\
Valpara\'iso, Chile,}
\affiliation{Centro de Astrof\'isica de Valpara\'iso, \\Gran Breta\~na 1111, Playa Ancha,
\\Valpara\'{\i}so, Chile.}

\author{ Yerko V\'asquez }
\email{yerko.vasquez@ufrontera.cl}
\affiliation{\it Departamento de Ciencias F\'isicas, Facultad de Ingenier\'ia,
Ciencias y Administraci\'on, Universidad de La Frontera, Avenida Francisco
Salazar 01145, Casilla 54-D,\\ Temuco, Chile.}

\date{\today}

\begin{abstract}
In this paper we studied the behavior of radial
photons from the point of view of the
coordinate time in (asymptotically) Lifshitz space-times,
and we found a generalization to the result reported in previous works
by Cruz et. al.  [Eur. Phys. J. C {\bf 73}, 7 (2013)],
Olivares et. al. [Astrophys. Space Sci. 347, 83-89 (2013)],
and Olivares et. al. [arXiv: 1306.5285].
We demonstrate that all asymptotically Lifshitz space-times characterized by a lapse
funcion $f(r)$ which tends to one when $r\rightarrow \infty$, present the same behavior,
in the sense that an external observer will see that photons arrive at spatial infinity
in a finite coordinate time. Also, we show that radial photons in the proper system cannot
determine the presence of the black hole in the region $r_+<r<\infty$,
because the proper time results to be independent of the lapse function $f(r)$.
\end{abstract}

\pacs{04.20.Fy, 04.20.Jb, 04.40.Nr, 04.70.Bw}

\keywords{Black Holes;  Elliptic Functions.}

\maketitle

\tableofcontents


\section{Introduction}
\label{intro}
Lifshitz space-times have proven to be important from a holographic point of view.
They represent gravity duals of non-relativistic systems and they are of particular
interest in the studies of critical exponent theory and phase transitions
\cite{Kachru,Hartnoll:2009ns}. In these space-times the spatial
and the temporal coordinates scale in different ways,
reflecting the symmetries of boundary field theory.

Lifshitz vacua space-times in d-dimensions are represented by the line element

\begin{equation}
ds^{2}=-\frac{r^{2z}}{\ell^{2z}}\, dt^{2}+\frac{\ell^{2}}{r^{2}}\,dr^{2}+r^{2}d\vec{x}%
^{2} , \label{metric}
\end{equation}%
where $z$ is the dynamical exponent, $\ell$ is the only length
scale in the geometry and $\vec{x}$ stands for a spatial (d-2)-dimensional vector.

This metric is invariant under the following change of coordinates:

\begin{equation}
t\rightarrow \lambda ^{z}\,t,\quad x\rightarrow \lambda\, x ,\quad r\rightarrow r/\lambda.
\end{equation}%
In this way, here $z$ is understood as a measure of the
anisotropy between spatial and temporal coordinates.
For $z=1$ we recover the AdS metric in Poincar\'e coordinates.

Black hole solutions whose asymptotic behavior is given by metric (\ref{metric})
have been studied recently. Lifshitz black holes with a flat transverse section in
four dimensions with $z=2$ have been reported in \cite{Balasu,Taylor:2008tg},
a topological Lifshitz black hole with $z=2$ was found in \cite{Mann}, a
black hole with spherical horizon with $z=4$ was found in \cite{Bertoldi:2009vn},
a three-dimensional Lifshitz black hole with $z=3$ was reported in \cite{AyonBeato:2009nh},
black holes of Lovelock theory can be found in \cite{Dehghani:2010kd}.

Recently, in \cite{COV13,germancito,felipito} the authors
showed that radial photons in those Lifshitz backgrounds can reach the asymptotic
region in a finite coordinate time, as seen by an external observer. Motivated by
these results we analyzed Lifshitz space-times with an arbitrary dynamical exponent z,
and showed that this behavior still holds.

The paper is organized as follows. In section II we obtain the fundamental
equations for massless particles in the (asymptotically) Lifshitz space-time,
then we analyze the radial  motion in terms of the coordinate time.
Also, we analyze different classes of lapse function that appear in
the literature. In section III we analyze the radial motion of photons
in terms of the proper time.
Finally, in section IV we discuss our results and
make conclusions.


\section{Generic Lifshitz black hole space-time}
\label{GLBHS}

The metric of a generic Lifshitz space-times can be written as
\begin{equation}\label{g1}
  ds^2=-\frac{r^{2z}}{\ell^{2z}}f(r)\,dt^2+\frac{\ell^2}{r^2}\frac{dr^2}{f(r)}+\frac{r^2}{\ell^2}d\overrightarrow{x}^2
\end{equation}
where $\overrightarrow{x}$ is a (d-2)-dimensional vector, and
the function $f(r)$ depends only on the radial coordinate.
Furthermore, this function takes the value $f(r) = 1$ when $r \rightarrow \infty$, which means that
the metrics are asymptotically Lifshitz.
So, the Lagrangian associated with the metric (\ref{g1}) for radial photons is given by \cite{COV13,COV}
\begin{equation}\label{g2}
  2\mathcal{L}=-\frac{r^{2z}}{\ell^{2z}}f(r)\,\dot t^2+\frac{\ell^2}{r^2}\frac{\dot r^2}{f(r)}=0,
\end{equation}
which is independent of the space-time dimension, and the
dot represents a derivative
with respect to an affine parameter along the trajectory.

Thus, we can immediately obtain the following quadrature
\begin{equation}\label{g3}
  \frac{dr}{dt}=\pm \frac{r^{z+1}}{\ell^{z+1}}f(r),
\end{equation}
which will be solved by explicitly giving the radial function $f(r)$.
Therefore, assuming that photons are placed in $r=R_0$ when $t=0$, we can write
the above equation as
\begin{equation}\label{g4}
  t=\pm\, \ell^{z+1} \int_{R_0}^{r}\frac{d r'}{r'^{z+1}f(r')}.
\end{equation}
In order to obtain a full description, we will study
different space-times characterized by the radial function $f(r)$ separately.

\subsection{Lifshitz Vacua Space-time}
\begin{figure}[!h]
  \begin{center}
    \includegraphics[width=85mm]{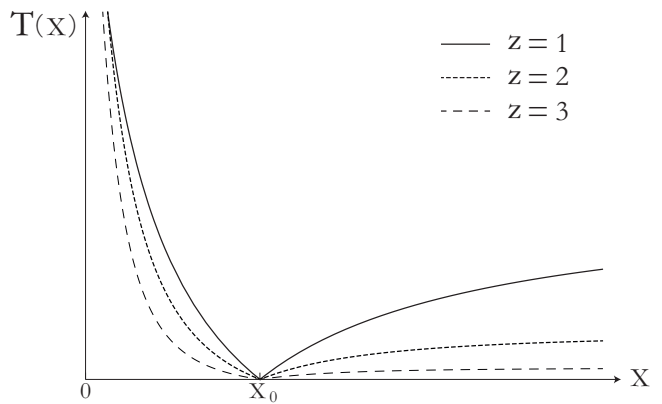}
  \end{center}
  \caption{Plot for the dimensionless coordinate time, $T \equiv t/\ell$, of photons generated by
  the lapse function $f(X)=1$ given by Eq. (\ref{g6}), where $X\equiv r/\ell$ and
  $X_0\equiv R_0/\ell$. Here we presents
  several curves for different values of the
  dynamical exponent $z$.
  This shows that, as seen by a system external to photons, they will fall
asymptotically to origin $X = 0$.
On the other hand, this external observer will see
that photons arrive at the spatial infinite
in a finite coordinate time, $T_1$, given by Eq. (\ref{g7}).}
  \label{f1}
\end{figure}

This case represents a space-time without black holes, i. e.,
the case with $f(r)=1$ \cite{Balasu}. Thus, Eq. (\ref{g4}) is written as
\begin{equation}\label{g5}
  t(r)=\pm\, \ell^{z+1} \int_{R_0}^{r}\frac{d r'}{r'^{z+1}},
\end{equation}
so, a straightforward integration yields to
\begin{equation}\label{g6}
  T(X)=\pm\, \left[\left(\frac{1}{X_0}\right)^z-\left(\frac{1}{X}\right)^z\right],
\end{equation}
where $T=t/\ell$ and $X=r/\ell$ are dimensionless variables, and $X_0=R_0/\ell$.
Thus, an external observer sees that photons moving from $X=X_0$ to $X=0$
need an infinite coordinate time to do so, whereas photons moving from
$X_0$ to the spatial infinite needs a finite time, $T_1$, given by
\begin{equation}\label{g7}
  T_1=\lim_{X\rightarrow \infty }T\left( X\right)=\left(\frac{1}{X_0}\right)^z.
\end{equation}
In FIG. \ref{f1} we plot the behavior
of the dimensionless function given by Eq. (\ref{g6}).

We know from the above analysis that for Lifshitz vacua space-times, the
integral

\begin{equation}
t_1= \ell^{z+1}\,\int_{R_{0}}^{\infty }\frac{dr}{r^{z+1}},
\label{Integral}
\end{equation}
converges. Now, we would like to know if the following integral converges

\begin{equation}
t_1= \ell^{z+1}\,\int_{R_{0}}^{\infty }\frac{dr}{r^{z+1}f\left(
r\right) },  \label{Integralsecond}
\end{equation}
for a generic $f\left( r\right) $, with $f\left( r\right) \rightarrow 1$
when $r\rightarrow \infty $.

In order to determine this, we apply the limit comparison test for improper
integrals. First, defining $f_{1}\left( r\right) =\frac{1}{r^{z+1}}$ and $%
f_{2}\left( r\right) =\frac{1}{r^{z+1}f\left( r\right) }$, both positive
definite in the interval $[R_{0},\infty )$.

Now, taking the limit we obtain
\[
a=\lim_{r\rightarrow \infty }\frac{f_{1}\left( r\right) }{f_{2}\left(
r\right) }=\lim_{r\rightarrow \infty }f\left( r\right) =1.
\]
This limit is $a=1$ for asymptotically Lifshitz black holes; therefore, in
accordance with the criteria for integral convergence, the integral (\ref%
{Integralsecond}) is convergent. This means that it takes a finite
coordinate time to reach the asymptotic region of radial photons. In the
next sections we will consider generic functions $f\left( r\right) $ and evaluate
explicity the integrals (\ref{Integralsecond}) for arbitrary $z>0$.

\subsection{Asymptotically Lifshitz black hole I}
\begin{figure}[!h]
  \begin{center}
    \includegraphics[width=85mm]{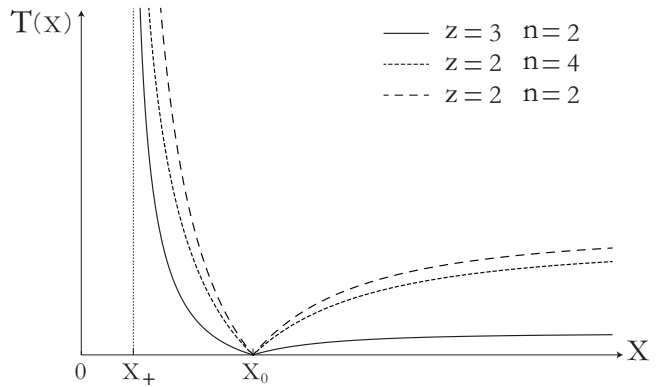}
  \end{center}
  \caption{Plot for the dimensionless coordinate time, $T\equiv t/\ell$, of photons generated by
  the lapse function $f(X)=1-X_+^n/X^n$ given by Eq. (\ref{g10}), where $X\equiv r/\ell$, $X_+\equiv r_+/\ell$ and
  $X_0\equiv R_0/\ell$. Here we present
  several curves for different values of the
  dynamical exponent $z$ and the parameter $n$.
  This shows that, as seen by a system external to photons, they will fall
  asymptotically to event horizon $X = X_+$.
  On the other hand, this external observer will see
  that photons arrive at the spatial infinite
  in a finite coordinate time, $T_1$, given by Eq. (\ref{g11}).}
  \label{f2}
\end{figure}

Here we will consider the family of asymptotically Lifshitz black holes whose
radial function have the form
\begin{equation}\label{g8}
  f(r)=1-\frac{r_+^n}{r^n}.
\end{equation}
Here $r_+$ is the so-called event horizon of the Lifshitz black hole, and $n$ is a real number which may
depend (or not) on the dynamical exponent $z$.
Functions of this kind can be found, for example, in \cite{AyonBeato:2009nh,eloy10,Balasu,Mann}.
Therefore, we can rewrite Eq. (\ref{g4}) as
\begin{equation}\label{g9}
  t(r)=\pm\, \ell^{z+1} \int_{R_0}^{r}\frac{d r'}{r'^{z-(n-1)}\,\left(r'^n-r_+^n\right)},
\end{equation}
in which case, a straightforward integration
leads us to the solution
\begin{equation}\label{g10}
  T(X)=\pm\,\frac{1}{z}\left[\Omega(X_0; X_+, z, n) -\Omega(X; X_+, z, n)\right],
\end{equation}
where the function $\Omega$ is given explicitly by
\begin{equation}\label{g101}
  \Omega(y; X_+, z, n)=\frac{1}{y^z}\,_2F_1\left(1, \frac{z}{n}, \frac{n+z}{n}, \frac{X_+^n}{y^n}\right),
\end{equation}
and $X_+\equiv r_+/\ell$. In Fig. \ref{f2} we plot the behavior
of the dimensionless function given by Eq. (\ref{g10}).

It can be ascertained here that an external observer will see that
photons take an infinite coordinate time to arrive at the
event horizon, which is a common fact with Einstein's
space-times (S, SdS, SAdS, etc.).
Moreover, again we obtain the situation described in the previous example,
i. e., there is a finite coordinate time in which
photons arrive at the spatial infinite given by
\begin{equation}\label{g11}
T_1=\lim_{X\rightarrow \infty }T\left( X\right)=
\frac{\,_2F_1\left(1, \frac{z}{n}, \frac{n+z}{n}, \frac{X_+^n}{X_0^n}\right)}{z\,X_0^z}.
\end{equation}

Recently, this behavior has been reported in \cite{COV13} for
$z=3$ and $n=2$, in \cite{germancito} for $z=2$ and $n=2$, and in
\cite{felipito} for $z=2$ and $n=4$.
\medskip

\subsection{Asymptotically Lifshitz black hole II}
In addition to the case above,
let us consider the following function:
\begin{equation}\label{g12}
  f(r)=\left(1-\frac{r_+^n}{r^n}\right)\,\left(1+\frac{g}{r^n}\right),
\end{equation}
where $r_+$ is the event horizon and $g$ is
any real constant. This kind of lapse function can be founded, for example, in \cite{Bertoldi:2009vn}. Therefore Eq. (\ref{g4})
can be written as
\begin{equation}\label{g13}
  t(r)=\pm\, \ell^{z+1} \int_{R_0}^{r}\frac{d r'}{r'^{z-(2n-1)}\,\left(r'^n-r_+^n\right)\,\left(r'^n+g\right)},
\end{equation}
and its dimensionless solution is given by
\begin{equation}\label{g14}
  T(X)=\pm\,\frac{\Psi(X_0, X_+, z, n, \tilde g)-\Psi(X, X_+, z, n, \tilde g)}{z\,(\tilde g+X_+^n)},
\end{equation}
 where the function $\Psi$ is given explicitly by
\begin{widetext}
\begin{equation}\label{g15}
  \Psi(y, X_+, z, n, \tilde g)=\frac{X_+^n}{y^z}\,_2F_1\left(1, \frac{z}{n}, \frac{n+z}{n}, \frac{X_+^n}{y^n}\right)
  + \frac{\tilde g}{y^z}\,_2F_1\left(1, \frac{z}{n}, \frac{n+z}{n}, -\frac{\tilde g}{y^n}\right),
\end{equation}
\begin{figure}[!h]
  \begin{center}
    \includegraphics[width=180mm]{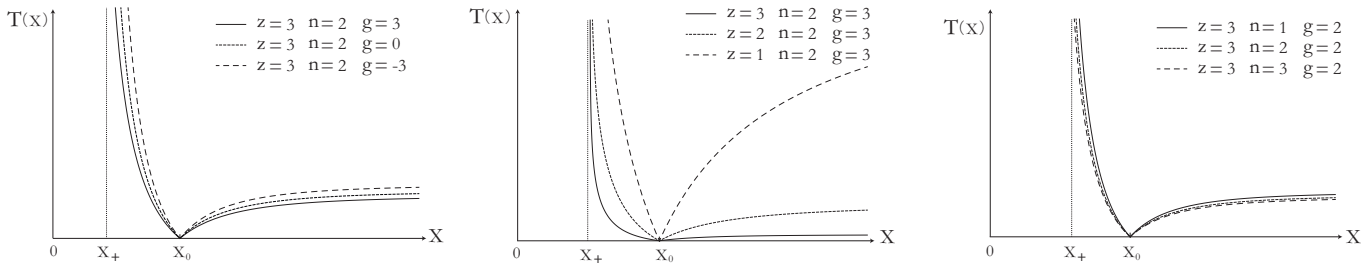}
  \end{center}
  \caption{Plot for the dimensioneless coordinate time,  $T\equiv t/\ell$, of photons generated by
  the lapse function $f(X)=(1-X_+^n/X^n)(1+\tilde g/X^n)$ given by Eq. (\ref{g14}), where $X\equiv r/\ell$,
  $X_+\equiv r_+/\ell$, $X_0\equiv R_0/\ell$ and $\tilde g \equiv g/\ell^n$. LEFT PANEL:
  Curves with $z=3$ and $n=2$ for different values of the constant $g$; MIDDLE PANEL:
  Curves with $n=2$ and $g=3$ for different values of the dynamical exponent, $z$; RIGHT PANEL:
  Curves with $g=2$ and $z=3$ for different values of the parameter $n$.
  This shows that, as seen by a system external to photons, they will fall
  asymptotically to event horizon $X = X_+$.
  On the other hand, this external observer will see
  that photons arrive at the spatial infinite
  in a finite coordinate time, $T_1$, given by Eq. (\ref{g14}).}
  \label{f3}
\end{figure}
\end{widetext}
and $\tilde g\equiv g/\ell^n$. In FIG. \ref{f3} we plot the behavior
of the dimensionless function given by Eq. (\ref{g14}).
Again, we found the same qualitative behavior as preceding cases. An external observer will see that
photons take an infinite coordinate time to arrive at event horizon, while they arrive at
the  spatial infinite in a finite coordinate time given by
\begin{equation}\label{g16}
  T_1=\lim_{X\rightarrow \infty }T\left( X\right)=
  \frac{\Psi(X_0, X_+, z, n, \tilde g)}{z\,(\tilde g+X_+^n)}.
\end{equation}

\section{A note on the proper system}
Since the metric of a Lifshitz space-time is static,
the Lagrangian associated is independent
of the temporal coordinate $t$, and thus, its corresponding
canonical conjugate momentum is a conserved quantity.
So, considering the Lagrangian (\ref{g2}) we obtain
\begin{equation}\label{p1}
  \Pi_t\equiv \frac{\partial \mathcal{L}}{\partial \dot t}=
  -\frac{r^{2z}}{\ell^{2z}}\,f(r)\,\dot t =-\sqrt{E}.
\end{equation}
Therefore, by combining Eq. (\ref{g3}) with Eq. (\ref{p1})
we obtain the quadrature
\begin{equation}\label{p2}
  \frac{dr}{d\tau}=\pm \frac{\ell^{z-1}}{r^{z-1}}\,\sqrt{E},
\end{equation}
and its dimensionless solution is given by
\begin{equation}\label{p3}
  \Theta(X)=\pm\frac{X^z-X_0^z}{\sqrt{E}},
\end{equation}
where $\Theta \equiv \tau/\ell$ is the dimensionless proper time.
The first observation of the solution (\ref{p3})
is that the proper system no detects
the shape of the lapse function $f(r)$, i. e.,
if photons are in the region $r_+<r<\infty$ ($X_+<X<\infty$),
then they cannot determine the presence of black hole,
which is in agreement with  General Relativity,
moreover,  the Schwarzschild case is recuperated when $z=1$.
In this sense, massless particles presents the same
asymptotic behavior as in Einstein's space-times:
\begin{enumerate}
 \item They cross the event horizon in a finite
 (dimensionless) proper time $\Theta_+$
 given by
 \begin{equation}\label{p4}
   \Theta_+\equiv\Theta(X_+)=\frac{X_0^z-X_+^z}{\sqrt{E}},
 \end{equation}
 and, eventually, arrives at the origin in a finite (dimensionless) proper time given  by
 \begin{equation}\label{p5}
   \Theta_0\equiv \Theta(X=0)=\frac{X_0^z}{\sqrt{E}}.
 \end{equation}
 \item They requires an infinite (dimensionless) proper time to arrive at the spatial infinite, i. e.,
 $\tau\rightarrow\infty$ when $r\rightarrow\infty$.
\end{enumerate}
\begin{figure}[!h]
  \begin{center}
    \includegraphics[width=85mm]{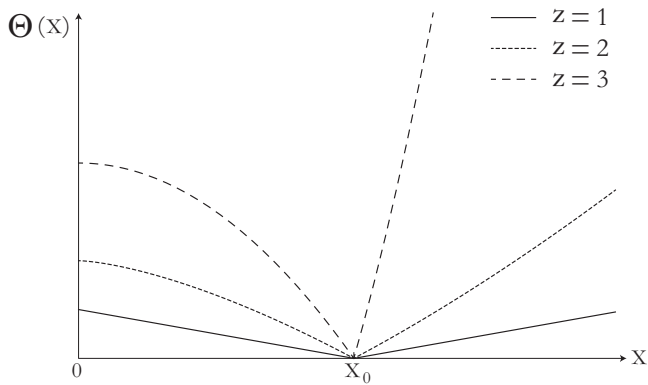}
  \end{center}
  \caption{Plot for the dimensionless proper time, $\Theta \equiv \tau/\ell$, as a function
  of the dimensionless radial coordinate $X\equiv r/\ell$, where
  $X_0\equiv R_0/\ell$. Here we present
  several curves for different values of the
  dynamical exponent $z$ with the same value of the constant of motion $E$.
  Notice that the Schwarzschild case is recuperated when $z=1$.}
  \label{f4}
\end{figure}
\section{Summary}
\label{Sum}
We obtained a generalization for the behavior
of massless particles in a Lifshitz space-time reported
in previous works \cite{COV13,germancito,felipito}. This generalization
is independent of the dynamical exponent $z$, and the dimension
of space-time. We found that space-time looks similar
to space-times of general relativity in the sense that an
external observer measures an infinite coordinate time
to photons traveling to the event horizon (or to the origin,
if a Lifshitz vacua space-time is considered).
On the other hand, by using of a simple criterion of convergence,
we found a general condition over the measure of coordinate time
employed by photons to arrive at spatial infinite. Our result is
that an external observer will see that  photons arrive at the spatial
infinite in a finite coordinate time, and the condition is that the
lapse function, $f(r)$, tends to one when $r\rightarrow \infty$.
Obviously, this condition always satisfies the asymptotic Lifshitz
space-times, and therefore, our novel result becomes general for all
space-times that satisfy the above condition. Additionally, we obtained that,
from the point of view of the proper system, massless particles present the same behavior
as Einstein's space-times, because the proper time is independent of the lapse
function $f(r)$, therefore, the photon cannot determine the presence of the black hole
in the region $r_+<r<\infty$.

\begin{acknowledgments}
Y. V. is supported by FONDECYT grant 11121148;\\
J. R. V. thanks the UFRO for their hospitality.
\end{acknowledgments}



\begin{thebibliography}{9}
\bibitem{Kachru}
Kachru S., Liu X. and Mulligan M.:
Gravity Duals of Lifshitz-like Fixed Points.
Phys. Rev. \textbf{D} 78, 106005 (2008) [arXiv: 0808.1725].

\bibitem{Hartnoll:2009ns}
Hartnoll S.~A., Polchinski  J., Silverstein  E. and Tong D.:
Towards strange metallic holography.
JHEP \textbf{1004}, 120 (2010).

\bibitem{Balasu}
Balasubramanian K. and McGreevy J.:
An analytic Lifshitz black hole.
Phys. Rev. {\bf D} 80, 104039 (2009) [arXiv: 0909.0263].

\bibitem{Taylor:2008tg}
Taylor M.:
Non-relativistic holography.
[arXiv: 0812.0530].

\bibitem{Mann}
Mann R. B.:
Lifshitz topological black holes.
JHEP \textbf{06}, 075 (2009). [arXiv: 0905.1136].

\bibitem{Bertoldi:2009vn}
Bertoldi  G.,~Burrington B.~A. and Peet  A.:
Black Holes in asymptotically Lifshitz space-times with arbitrary critical exponent.
Phys. Rev. {\bf D} 80, 126003 (2009)
[arXiv: 0905.3183].


\bibitem{AyonBeato:2009nh}
Ayon-Beato E., Garbarz A., Giribet G. and Hassaine M.:
Lifshitz Black Hole in Three Dimensions.
 Phys. Rev.  {\bf D} 80, 104029 (2009) [arXiv: 0909.1347].

\bibitem{Dehghani:2010kd}
Dehghani M.~H. and Mann R.~B.:
Lovelock-Lifshitz Black Holes.
JHEP \textbf{1007}, 019 (2010) [arXiv: 1004.4397].

\bibitem{COV13}
Cruz N., Olivares M. and Villanueva J. R.:
Geodesic structure of the Lifshitz Black Hole in 2+1 dimensions.
Eur. Phys. J. C {\bf 73}, 7 (2013) [arXiv: 1305.2133].

\bibitem{germancito}
Olivares M., Rojas G., V\'asquez Y. and Villanueva J. R.:
Particles motion on topological Lifshitz black holes in 3+1 dimensions.
Astrophys. Space Sci. 347, 83-89 (2013) [arXiv: 1304.4297].

\bibitem{felipito}
Olivares M., V\'asquez Y., and Villanueva J. R., Moncada F.:
Motion of particles on a $z=2$ Lifshitz black hole in 3+1 dimensions.
(2013) [arXiv: 1306.5285].

\bibitem{COV}
Cruz N., Olivares M. and Villanueva J. R.:
The geodesic structure of the Schwarzschild anti-de Sitter Black Hole.
Class. Quantum Grav. \textbf{22}, 1167-1190 (2005) [arXiv: 0408016].

\bibitem{eloy10}
Ayon-Beato E., Garbarz A., Giribet G. and Hassaine M.:
Analytic Lifshitz black holes in higher dimensions.
JHEP {\bf 1004}, 030 (2010)
[arXiv: 1001.2361]







\end{thebibliography}
\end{document}